\documentclass[12pt]{article}

\usepackage{scicite}

\usepackage{times}

\usepackage{graphicx}

\title{Entangled, Spin-polarized Excitons from Singlet Fission in a Rigid Dimer}

\author
{Ryan D. Dill$^{1}\dagger$, Kori E. Smyser$^{1}\dagger$, Niels H. Damrauer$^{1,2\ast}$, Joel D. Eaves$^{1,2\ast}$\\
\\
\normalsize{$^{1}$Department of Chemistry, University of Colorado Boulder;}\\
\normalsize{Boulder, CO, 80309, USA.}\\
\\
\normalsize{$^{2}$Renewable and Sustainable Energy Institute (RASEI), University of Colorado Boulder;}\\
\normalsize{Boulder, CO, 80309, USA.}\\
\\
\normalsize{$\dagger$These authors contributed equally to this work.}
\\\normalsize{$^\ast$Corresponding authors. Emails: niels.damrauer@colorado.edu (NHD);} \\\normalsize{joel.eaves@colorado.edu (JDE)}
}

\date{}

\begin{document} 


\baselineskip24pt


\maketitle 



\begin{abstract}
Singlet fission, a process that splits a singlet exciton into a biexciton, has promise in quantum information. We report time-resolved electron paramagnetic resonance measurements on a molecule, TIPS-BP1$'$, designed to exhibit strongly state-selective relaxation to specific magnetic spin sublevels. The resulting optically pumped ``spin polarization'' is a nearly pure initial state from the ensemble. The long-lived spin coherences modulate the signal intrinsically, allowing a new measurement scheme that substantially removes noise and uncertainty in the magnetic resonance spectra. A nonadiabatic transition theory with a minimal number of spectroscopic parameters allows the quantitative assignment and interpretation of the spectra. The rigid, covalently bound dimer, TIPS-BP1$'$, supports persistent spin coherences at temperatures far higher than those used in conventional quantum hardware.

\subsubsection*{One-Sentence summary:}
Theory guides identification of a singlet fission dimer to selectively form long-lived EPR-active states at high temperature.
\end{abstract}

\flushbottom
\clearpage

Quantum information promises advances in science and computing not seen since the revolutions in classical computing that have unfolded over the last 80 years \cite{DiVincenzo.2000}. But unlike classical computing, where the solid-state transistor has become ubiquitous, we remain in the discovery phase for quantum materials. Quantum logic uses fragile non-equilibrium quantum states built upon qubits that irreversibly decay to Boltzmann equilibrium. In strong-field experiments, microwave or radio frequencies  manipulate the qubits to perform operations \cite{Vandersypen.2001}. Because the resonant frequencies are much smaller than the thermal energy at room temperature, without extreme cooling or other means of control, a significant population in the excited state generates thermal uncertainty in the initial state of the wavefunction \cite{Warren.1997}. This ``tyranny of temperature" makes quantum circuits classical for temperatures above a few kelvin \cite{Smyser.2020}.

Removing the uncertainty in the initial condition of the wavefunction solves the so-called ``state-initialization problem," a requirement for quantum computation that DiVincenzo articulated more than twenty years ago \cite{DiVincenzo.2000}. For example, in color centers, like nitrogen-vacancy centers in diamond, a weak-field optical excitation initializes the system into a non-equilibrium state---a magnetic sublevel---where strong-field magnetic resonance pulses perform gate operations \cite{Dutt.2007}. But controlling the placement of defects in crystals is challenging, which makes scaling the number of qubits in these materials a formidable hurdle. Recent molecular analogs to the color centers suggest that a bottom-up approach from synthetic chemistry might ultimately lead to more scalable architectures \cite{Bayliss.2020}. Like many other quantum materials, however, the molecules only exhibit quantum function near liquid helium temperatures.

In this article, we take a bottom-up approach and initialize qubits from spin-polarized biexcitons in a structurally well-defined organic molecule. Molecular symmetry is exploited to produce selection rules that solve the state-initialization problem using singlet fission (SF), a photophysical process that produces a highly spin-entangled biexciton state $^1\textrm{TT}$ with singlet multiplicity. In an earlier publication, Smyser and Eaves predicted that a molecular dimer, with its chromophores oriented so that their principal axes are parallel, can convert $^1\textrm{TT}$ into a strongly spin-polarized quintet biexciton state, $^5\textrm{TT}$ \cite{Smyser.2020}. The principal axes of the chromophores in the dimer studied here, TIPS-BP1$'$, are not \emph{all} parallel, but their $y$-axes are. As a result, TIPS-BP1$'$ exhibits an intense spin polarization manifest in time-resolved electron paramagnetic resonance (trEPR) spectra. The symmetry of the molecule (C\textsubscript{2v}) and the rigidity of its structure ensure that parasitic photoproducts from SF are undetectable and that electron spin coherence remains for microseconds---even in a glassy phase at liquid nitrogen temperatures. We both predict and detect the spin sublevels that immediately follow SF at a level of accuracy that is unprecedented in the literature.

The several possible biexciton species $^{2S+1}\textrm{TT}_M$ that differ in their overall spin $S$ and degree of entanglement are not directly distinguished by transient absorption spectroscopy,\cite{Gilligan.2019} so we turn to trEPR to resolve them. The experiment starts the SF process with an optical pulse and then uses EPR to monitor the time-evolution of the products. TrEPR signatures of TIPS-BP1$'$ in mTHF glass (75 K, 640 nm pump wavelength) emerge over a few hundred nanoseconds following photoexcitation (Fig.~1A). This timescale is consistent with the decay of $^1\textrm{TT}$ (Fig.~S1) and is impulsive on the timescale of the trEPR measurement (10 $\mu$s) \cite{Gilligan.2019}. Four sharp features, from 338-359 mT, dominate the trEPR spectra for all observable times. They form concomitantly and exhibit underdamped Rabi oscillations that beat at the nutation frequency (Fig.~1A, inset). These oscillations have not been reported in trEPR data for any system undergoing SF but have been observed for triplets where relaxation processes are slow \cite{Schroder.2021}.

General trEPR trends in the SF literature include broad and congested spectra, with substantial interconversion between EPR-active states \cite{Papadopoulos.2019,Basel.2018,Lubert-Perquel.2018,Nagashima.2018,Sakai.2018,Tayebjee.2017,Weiss.2017}. By contrast, our spectra---aside from the oscillations---do not show substantial time evolution.  They are also highly structured and symmetrical. The EPR spectra in Fig.~1A are narrow, with intensity spanning 20 mT. The intersystem crossing triplet spectrum for the monomer TIPS-Pc (Fig.~S3), in comparison, spans 84 mT. The relatively narrow width of the TIPS-BP1$'$ spectra suggests that the signal originates from $^5\textrm{TT}$ \cite{Benk.1981}.

Nutation frequencies depend on $S$ and $M$, so they can, in principle, inform on the spin species and sublevels produced after SF \cite{Astashkin.1990}. In the SF literature, they are commonly determined with pulsed EPR at only a few values of the static magnetic field, $B_0$ \cite{Basel.2018,Lubert-Perquel.2018,Nagashima.2018,Sakai.2018,Tayebjee.2017,Weiss.2017}. Compared to pulsed nutation experiments, trEPR has a dramatic multiplex advantage---an entire time trace is collected simultaneously (Fig.~1). However, pulsed techniques with high microwave powers are necessary for most SF systems since rapid dephasing and population transfer overdamp the low-frequency nutation oscillations in trEPR \cite{Stoll.1998}. In TIPS-BP1$'$, by contrast, the presence of Rabi oscillations at a dominant frequency in the trEPR data implies that there is a state-selective population formed rapidly on the timescale of the oscillation period, whose decoherence time is longer than a microsecond ($T_2 \approx 1.4~\mu$s, Fig.~S4).

In trEPR, Rabi oscillations for inhomogeneously broadened transitions decay as an exponentially damped Bessel function \cite{Furrer.1980}. Thus, the Hankel transform, which projects the signal onto the Bessel functions, substantially enhances frequency resolution relative to the Fourier transform (Fig.~1B, inset, and S5). Our method shares many similarities with lock-in detection, but rather than externally modulating the signal, the method ``locks in'' at the sharply peaked dominant nutation frequency $\omega_N$ to separate low-frequency components from the oscillating signal (Fig.~1B). The ``Hankel spectrum'' is the integrated intensity along the frequency axis within a prescribed bandwidth (Fig.~1B and S6). It isolates the signal that nutates at $\omega_N$---the majority component of the EPR data (Fig.~2, black lines).

A signal oscillating at a dominant nutation frequency might result from a state-selective relaxation process; from $^1\textrm{TT}$ into a few specific $^5\textrm{TT}_M$ sublevels, and such precise state-selectivity can solve the state-initialization problem in quantum information. But to determine the extent of state-selectivity in a molecule, an accurate interpretation of the EPR spectrum is essential. Some have adapted Merrifield's theory \cite{Johnson.1969} for triplet-triplet annihilation, to compute the $\textrm{TT}$ populations that the EPR experiment probes \cite{Tayebjee.2017,Weiss.2017,supplementary.materials}. Therein, when the inter-chromophore exchange interaction $J$ is zero the resulting spectrum only comes from $M=0\rightarrow{M=\pm1}$ transitions, so we refer to it as the `$\textrm{Q}_0$' model. But the $\textrm{Q}_0$ model is inappropriate for strongly-coupled dimers that directly populate $^5\textrm{TT}$, so it does not reproduce the spectrum of TIPS-BP1$'$ (Fig.~2A). Without a theory to determine the populations, they become fitting parameters \cite{Nagashima.2018}. In the dense and broad spectra typical of EPR data for SF, these additional parameters lead to uncertainty and overfitting that complicates the interpretation of the spectra.

To overcome this problem, we \emph{compute} the populations of the initial $^5\textrm{TT}$ sublevels with our nonadiabatic transition theory by extending the $JDE$ model \cite{Smyser.2020,supplementary.materials}. Unlike the $\textrm{Q}_0$ model, our model predicts that both the resonance field values \emph{and} the populations depend on a molecule's orientation relative to $\vec{B_0}$. Nuclear distortions cause rare, but large, fluctuations in $J$ that induce transitions from $^1\textrm{TT}$ to quintet sublevels---selection rules forbid relaxation from $^1\textrm{TT}$ to $^3\textrm{TT}$. The EPR experiment immediately following SF probes the ``prompt'' EPR spectrum and, for TIPS-BP1$'$, it comes entirely from transitions within the quintet manifold.

Because the experiment probes all orientations at once, it eliminates the uncertainties of the molecular orientation relative to $\vec{B_0}$ that one encounters in crystals \cite{Lubert-Perquel.2018}. Using the $JDE$ model, we first calculate the populations and EPR spectrum for a molecule with a given orientation relative to the lab frame (Fig.~3). The ensemble-averaged EPR spectrum is then a sum over all such spectra, for orientations drawn from the surface of a tessellated sphere. In this work, there are only three spectroscopic fit parameters: the ``zero-field'' parameters, $D$ and $X$ \cite{supplementary.materials}, and the dihedral angle between the chromophores, $\beta$ (Fig.~3B). Physically, both $D$ and $X$ characterize the strength of the anisotropic spin-spin interactions; $D$ is intra-chromophore, and $X$ is inter-chromophore.

The zero-field hamiltonian mixes the $^5\textrm{TT}_M$ sublevels so that there are two choices for the initial states after relaxation from $^1\textrm{TT}$: the adiabatic states that diagonalize the quintet blocks of the hamiltonian or the diabatic Zeeman states $\{|^5\textrm{TT}_M\rangle \}$ \cite{supplementary.materials}. The adiabatic initial states (Fig.~2A) give a more accurate reproduction of the EPR spectrum. But, because they depend on molecular orientations, they are not intuitive when assigning the spectrum. The diabatic Zeeman states are, however, well-defined in the lab frame, independent of orientation, and do facilitate assignment. Because the mixing between the $^5\textrm{TT}_M$ sublevels is weak $(|D|\ll|B_0|)$, there are only small, quantitative differences between the spectra calculated with the diabatic and adiabatic bases (Fig.~S9).

Figure~2A shows the prompt EPR spectrum for TIPS-BP1$'$ (200-400 ns, Figs.~1A and S2) along with a calculation of it. The best fit values $D=1322$ MHz and $X=59$ MHz are consistent with those for pentacene derivatives and dimers, respectively \cite{Lubert-Perquel.2018}. The fit value of $\beta = 111.1^\circ$ is within 0.2\% of the calculated value from DFT simulations for the quintet in a model of TIPS-BP1$'$ ($110.9^\circ$, unrestricted-$\omega$-B97XD/6-31G(d)).

With the optimal set of spectroscopic parameters determined, the calculated EPR spectrum breaks down into two components from the diabatic $^5\textrm{TT}_M\leftrightarrow{^5\textrm{TT}_{M\pm1}}$ transitions. Figures~2B and 2C show the results. Our theory demonstrates that the Hankel transform isolates the signal from the $^5\textrm{TT}_{0}\leftrightarrow {^5\textrm{TT}_{\pm1}}$ transitions (Fig.~2B), and supports the assignment of the nutation frequency to this component. The residual spectrum (Figs.~2C, S7 and S8), the difference between the Hankel spectrum (Fig.~2B) and the total spectrum (Fig.~2A), agrees with the computed $^5\textrm{TT}_{\pm1} \leftrightarrow {^5\textrm{TT}_{\pm2}}$ spectral component quantitatively, in both amplitude and functional form.

It is only by accounting for the orientational dependence of the sublevel populations that we recover the spectrum from TIPS-BP1$'$. Figure~3A shows that the most intense features in the powder spectra are from transitions where the Zeeman field aligns with the dimer axes (Figs.~3B and 3C). Figure~3A also shows that while the $^5\textrm{TT}_0$ sublevel population is large for $B_0\parallel z$ and $\vec{B}_0\parallel y$, it is \emph{zero} for $\vec{B}_0\parallel x$. In the $\textrm{Q}_0$ model, by contrast, the $^5\textrm{TT}_0$ sublevel is the only $\textrm{TT}$ sublevel populated for any orientation---including $\vec{B}_0\parallel x$---leading to an over-representation of the $^5\textrm{TT}_0\rightarrow {^5\textrm{TT}_{\pm1}}$ transitions in the spectrum, and a poor resulting fit (Fig.~2A, gray). Indeed, if the $^5\textrm{TT}_0$ level were the only sublevel populated, the residual spectrum would be zero.

To engineer a piece of quantum hardware based on our system and observations, one would have to immobilize and align the molecules so that they all have a definite orientation with respect to the Zeeman field. Figures~3D and 3E show  the predicted spin polarization for a system of aligned TIPS-BP1$'$ dimers. Borrowing an idea from Shannon's classical information theory \cite{Kardar.2007}, we introduce the order parameter $\mathcal{I}$ to quantify the spin polarization achievable into \emph{any} $^5\textrm{TT}_M$ sublevel from $^1\textrm{TT}$ as a function of molecular orientation relative to the field (Fig.~3C), where $\mathcal{I} = 1 + \frac{1}{\log_2(5)}\sum_{M=-2}^{+2}p_M\log_2p_M$ (Fig.~3D). Like Shannon's information measure, $\mathcal{I}$ is zero when all $^5\textrm{TT}_M$ are equally populated and unity when only one level is occupied. Our work in ref.~\cite{Smyser.2020} recommends that the chromophores share a common set of axes. While the $x$ and $z$-axes of the chromophores are not parallel for TIPS-BP1$'$, the $y$-axes \emph{are}. As a result, the most intense spin polarization occurs when the Zeeman field aligns with the shared $y$-axis \cite{Lewis.2021}. The corresponding north and south poles of Fig.~3D exhibit the largest spin polarization, and Fig.~3E shows that the $^5\textrm{TT}_0$ sublevel is the one that gets polarized.

In molecular systems like those pioneered in nuclear spin resonance computing, scaling the number of coherent qubits is relatively straightforward \cite{Vandersypen.2001}. But the state-initialization problem has bedeviled that field \cite{Warren.1997}. TIPS-BP1$'$ is an example of a novel class of compounds that create entanglements between \emph{electron} spin states that remain coherent on timescales that are orders of magnitude longer ($\approx$ 1 $\mu$s) than the switching time for a gate operation ($\approx$ 1 ns), even in a powder spectrum. The quintet state, born under the selection rules of singlet fission, is a two-triplet spin-coherent excitation. The coherence entangles the triplets and increases the number of computational states from three to five---an elementary demonstration of scaling. Our results motivate efforts to orient TIPS-BP1$'$, and molecules like it, through crystallization or other means. SF in rigid molecular dimers solves the state initialization problem at temperatures far higher than the operating temperatures in contemporary quantum hardware.


\section*{Acknowledgments}
We thank Dr. Justin Johnson, Dr. Obadiah Reid, and Dr. Brandon Rugg for insightful discussions and Dr. Rugg for collection of the TIPS-Pentacene EPR spectrum. The dimer TIPS-BP1$'$ was synthesized previously by Ethan Miller in a collaboration of author NHD with Professor Tarek Sammakia of CU Boulder and funded by the National Science Foundation (CHE-166537). \textbf{Funding:} We acknowledge funding from the United States Department of Energy, Office of Basic Energy Sciences (ERW7404). This work made use of the EPR facility at the National Renewable Energy Laboratory, supported by the United States Department of Energy. This work also utilized resources from the University of Colorado Boulder Research Computing Group, which is supported by the National Science Foundation (awards ACI-1532235 and ACI-1532236), the University of Colorado Boulder, and Colorado State University. \textbf{Author contributions:} R.D.D. performed the measurements and K.E.S.  implemented the theory of K.E.S and J.D.E. Authors N.H.D and J.D.E. advised on all efforts. All authors contributed to the data analysis and manuscript preparation. \textbf{Data and materials availability:} All data are available in the main text or the supplementary materials.

\section*{Supplementary materials}
Materials and Methods\\
Supplementary Text\\
Figs. S1 to S9\\
Table S1\\
References \textit{(24-33)}

\clearpage

\section*{Figures}
\includegraphics[scale=1]{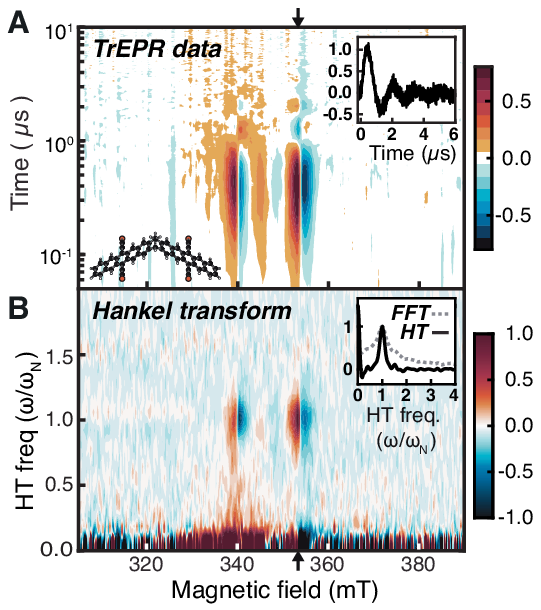}\\
\noindent {\bf Fig. 1. TrEPR data for TIPS-BP1$'$ demonstrates long spin-coherence times and strong spin polarization.}
({\bf A}) Contour plot of X-band trEPR data for TIPS-BP1$'$ (75 K and 640 nm excitation). Inset: Underdamped Rabi oscillations for a representative magnetic field value (353.4 mT, arrows). The decay closely follows a Bessel function which is expected for an orientationally distributed sample. ({\bf B}) The Hankel Transform (HT) then provides the nutation spectrum at each field point. Inset: HT of the transient shown in the inset of (A) peaks much more sharply than the comparable amplitude spectrum from the Fast Fourier Transform (FFT). This resolution enhancement facilitates extraction of the ``Hankel spectrum'' in Fig.~2B, which corresponds to $^5\textrm{TT}_0\leftrightarrow{^5\textrm{TT}_{\pm1}}$ transitions.
\clearpage

\includegraphics[scale=1]{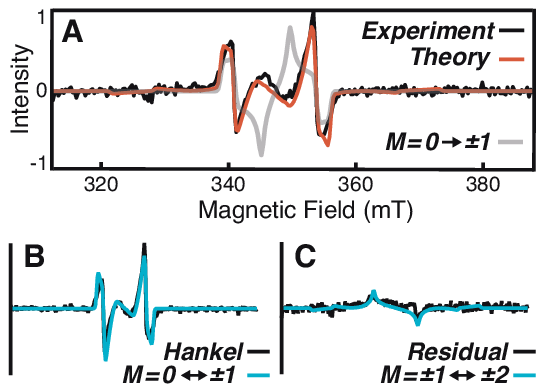}\\
\noindent {\bf Fig. 2. Data and calculated EPR spectrum for TIPS-BP1$'$.}
(\textbf{A}) The prompt trEPR spectrum for TIPS-BP1$'$ (black) is an average over 200-400 ns (Fig.~1A). The red line comes from the $JDE$ model with best-fit parameters $D = 1322$ MHz, $X = 59$ MHz, and $\beta = 111.1^\circ$ \cite{supplementary.materials}. Using the $\textrm{Q}_0$ model to predict initial populations (gray) does not reproduce the spectrum. (\textbf{B}) The Hankel spectrum (black) is the dominant signal and is replicated well by a calculated $^5\textrm{TT}_0\leftrightarrow{^5\textrm{TT}_{\pm1}}$ spectrum (blue, $JDE$ model). (\textbf{C}) The residual spectrum (black) is the difference between the full spectrum in (A) and the Hankel component in (B). It is reproduced (blue, $JDE$ model) with a calculated $^5\textrm{TT}_{\pm1}\leftrightarrow{^5\textrm{TT}_{\pm2}}$ spectrum. Any signal from triplets is undetectable. Relative amplitudes of the calculated spectra (blue) come from the $JDE$ model.
\clearpage

\includegraphics[scale=1]{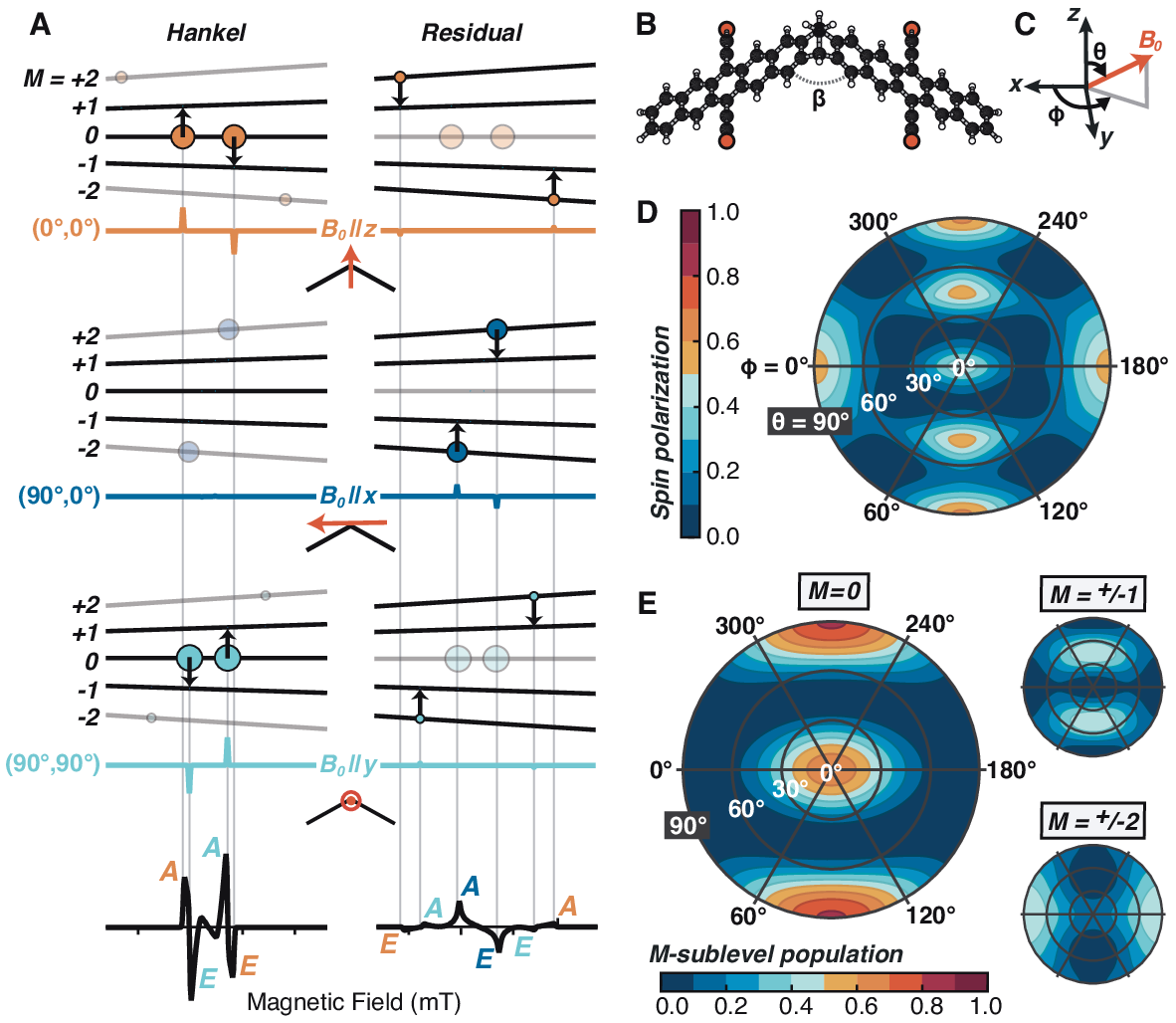}\\
\noindent {\bf Fig. 3. Theory predicts strong spin polarization for the rigid TIPS-BP1$'$ dimer. } (\textbf{A}) Fixed-orientation EPR spectra (colored lines) for $\vec{B}_0$ (red arrows) applied along cardinal dimer directions. Predictions for the Hankel spectrum ($^5\textrm{TT}_0\leftrightarrow {^5\textrm{TT}}_{\pm1}$) on left and residual spectrum ($^5\textrm{TT}_{\pm1}\leftrightarrow ^5\textrm{TT}_{\pm2}$) on right. Area of colored circles indicates $^5\textrm{TT}_M$-sublevel populations, and arrows show the direction of transitions ($A$ absorption or $E$ emission). Vertical lines correlate population assignments with features in the simulated powder spectra (black, below). (\textbf{B}) The two chromophores in the TIPS-BP1$'$ dimer are rigidly linked so that a single bridging angle $\beta$ defines the chromophore-chromophore orientation. (\textbf{C}) The polar and azimuthal angles $\theta$ and $\phi$ for the Zeeman field relative to the dimer cardinal axes. (\textbf{D}) Spin polarization $\mathcal{I} = 1 + \frac{1}{\log_2(5)}\sum_{M=-2}^{+2}p_M\log_2p_M$ for an ordered sample as a function of the dimer-field orientation. This range of $(\theta,\phi)$ considers all orientations with unique spectra. (\textbf{E}) $^5\textrm{TT}_M$ populations as a function of the dimer-field orientation. The $\pm M$-sublevels are predicted to be equally populated. The maximum population of $^5\textrm{TT}_0$ occurs at $\vec{B}_0\parallel y$, $(\theta,\phi) = (90^\circ,90^\circ)$.


\end{document}